\documentstyle[twoside,fleqn,espcrc2,epsf]{article}
\epsfverbosetrue
\title{\vspace{-.5in}\hbox{}\hfill{\normalsize ANL-HEP-CP-96-62}\\
\vspace{.3cm}
Manifestations of the axial anomaly in finite temperature QCD%
\thanks{Talk presented by J.-F. Laga\"e at LATTICE'96, St.~Louis MO,
4--8 June, 1996.}}
 
\author{J.B. Kogut\address {Department of Physics, University of Illinois,
                            1110 West Green Street, Urbana IL 61801, USA}%
\thanks{Supported in part by NSF grant NSF-PHY92-00148.},
J.-F. Laga\"e\address {HEP Division, Argonne National Laboratory, 
                       9700 South Cass Avenue, Argonne IL 60439, USA}%
\thanks{Supported by DOE contract W-31-109-ENG-38.} 
and D.K. Sinclair$\null^{{\rm b}\ddagger}$}

\begin{document}
 
\begin{abstract}
We compute the flavor singlet meson correlators and screening masses 
in quenched and $N_f=2$ QCD at $N_t=8$. The consequences of our results
for the realization of the $U_A(1)$ symmetry at finite T are discussed 
and an interpretation of our measurements in terms of the behaviour of 
the low lying fermionic modes is proposed.
\end{abstract}
 

\maketitle

Although the QCD chiral phase transition has been studied for many years, 
relatively little is known so far about the realization of the axial $U(1)$ 
symmetry in finite temperature QCD. The limiting cases are understood 
however: at very high temperature the $U_A(1)$ symmetry can be considered 
as ``effectively restored'' (since the effects associated with the axial 
anomaly are small in a dilute gas of instantons) whereas at zero temperature 
a sizeable explicit breaking of the $U_A(1)$ symmetry is essential to our 
understanding of meson spectroscopy. The question therefore arises as to how 
the transition between these two regimes is realized and how this relates to 
the restoration of the $SU(N_f)$ chiral symmetry.
In order to investigate this problem, we have computed mesonic screening 
correlators at various temperatures above and below the phase transition. 
Considering 2 flavors of valence quarks, we do this in 4 channels 
corresponding to the $\vec\pi$, $\sigma$, $\vec\delta$ and $\eta^\prime$. 
Chiral symmetry 
restoration would then induce degeneracies represented by the following 
grouping of states: $(\vec\pi , \sigma)$ and $( \vec\delta , \eta^\prime )$, 
while $U_A(1)$ restoration would imply: $( \sigma , \eta^\prime )$ 
and $( \vec\pi , \vec\delta )$. 

Our lattice computations are done using staggered quarks. The operators 
representing the $\vec\pi$, $\sigma$, $\vec\delta$ and $\eta^\prime$ are 
respectively $\gamma_5 \otimes \xi_5$, $ I \otimes I $, 
$ I \otimes \xi_5 $ and $ \gamma_5 \otimes I $. The first two being 
local operators and the last two 4-link operators. In order 
to minimize the effect of flavor symmetry breaking (which can be 
substantial at the values of $\beta$ currently used) it is desirable 
to limit comparisons to one of the two categories. This strategy allows 
a study of the ${ \vec\pi , \sigma , \vec\delta }$ system with minimal 
lattice artifacts if one replaces the $\vec\delta$ correlator by the 
connected part of the $\sigma$ correlator ($\sigma_{conn}$). This being 
said, the computations involving the $\eta^\prime$ operator are also quite 
important in exposing the underlying physics, since (through the 
Atyah-Singer index theorem) $\gamma_5 \otimes I$ serves as an indicator 
of topological activity in the QCD vacuum. All our correlators (connected 
and disconnected) were computed using a $U(1)$ noisy estimator following 
the techniques used by Kilcup et al. in zero temperature QCD \cite{KILCUP}. 
In addition to this and in order to help in the interpretation of our 
results we also computed the low lying spectrum of the Dirac operator 
( in practice the lowest 8 eigenvalues and associated eigenvectors ) 
on each of our configurations. This was done using the conjugate gradient 
algorithm investigated by Kalkreuter and Simma \cite{KALKR}. 
As we will show later, 
some of the correlators are entirely saturated by the few lowest fermionic 
modes, allowing further insight into the dynamics of finite temperature QCD. 

All our computations were done on the CRAY C-90 at NERSC on lattices of size $16^3 \times 8$. We have 
gathered results both for quenched and 2 flavor QCD. A single value of the 
quark mass has been studied so far: $ma=0.02$ for the quenched case and 
$ma=0.00625$ in 2 flavor QCD. In the later case, we use configurations 
generated by the HTMCGC collaboration \cite{00625}. 
The quenched computations are 
done at $\beta=5.8, 5.9, 6.0, 6.1$ and $6.2$ where we have respectively 
analyzed 100, 100, 100, 170 and 170 configurations. Similar numbers for 
the 2 flavor case are $\beta=$ 5.45 [160], 5.475 [240], 5.5 [160], 5.525 [160] 
and 5.55 [80]. The position of the crossover 
is found to be slightly higher than $\beta=6.0$ in the quenched case and 
slightly higher than $\beta=5.475$ in two flavor QCD. In figure 1, we plot 
our results for the $\vec\pi, \sigma$ and $\vec\delta$ screening lengths in 
2 flavor QCD as a function of $\beta$. The key feature of this plot is that 
the $\sigma$ becomes light close to the transition while the $\vec\delta$ 
remains 
heavy. This is in agreement with the observation that the peak in the scalar 
susceptibility originates in the disconnected part of the correlator 
\cite{KARSCH}. 
\begin{figure}[htb]
\vspace{-0.2cm}
\epsfxsize=7.5cm
\epsffile[70 70 550 550]{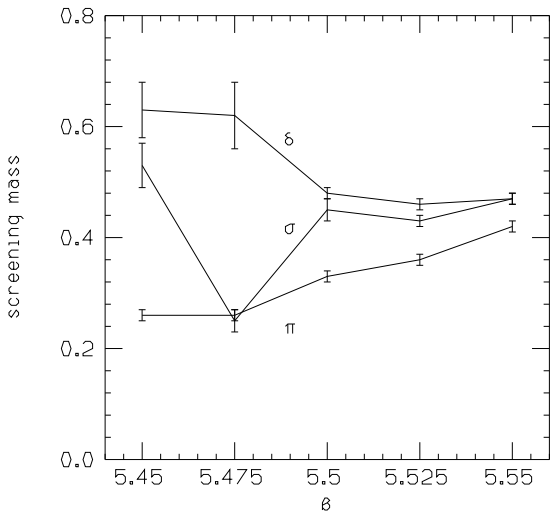}
\vspace{-1.3cm}
\caption{ Screening masses in two flavor QCD }
\vspace{-0.1cm}
\end{figure}
Above the phase transition, the $U_A(1)$ symmetry breaking 
($m_{\vec\delta} - m_{\vec\pi}$) appears to be of the same order as the 
explicit chiral symmetry breaking ($m_{\sigma} - m_{\vec\pi}$) at our 
current value of the quark mass ($ma=0.00625$). A careful extrapolation 
to the chiral limit will therefore be necessary in order to precisely 
sort out the situation in this region. This is currently being attempted 
by other groups \cite{CHRIST,DETAR,BOYD}, 
although they only compute the integrated correlators 
(susceptibilities) instead of the full correlators as we do here. 
Using the data already available at this stage, however, it is tempting 
to speculate that the $U_A(1)$ symmetry is not restored within the 
critical region. Using the HTMCGC data for the $\vec\delta$ correlator 
at $ma=0.0125$ \cite{0125} together with ours, it appears that $m_{\vec\delta}$ 
extrapolates to a non-zero number at all values of $\beta$. Assuming a 
second order transition however, $m_{\sigma}$ (and $m_{\vec\pi}$) will 
become very light close to the transition. Hence it appears that while 
$SU(2)_L \times SU(2)_R$ is being restored, $U_A(1)$ breaking is still 
present and only disappears at higher temperatures.

As mentioned earlier we have also computed the lowest 8 eigenvalues 
and associated eigenvectors on each of our configurations. Interestingly, 
we find that a spectral expansion of the quark propagator truncated to these 
low lying modes already captures a significant part of the physics. At or 
above the transition, for example, the pseudoscalar disconnected correlator 
is completely saturated by these low lying modes (similar computations below 
the phase transition are currently in progress). This suggests that a rather 
detailed description of the dynamics associated with the phase transition 
can be obtained.
In the continuum, we expect to find two kinds of low lying modes: exact 
zero modes (which must appear in number consistent with the index theorem: 
$n_+ - n_- = N_f Q_{top}$ ) and near zero modes (with no a priori connection 
to topology). Modes from the first category satisfy $r=1$ with 
$r=|<n|\gamma_5|n>|$, whereas those from the second necessarily have $r=0$. 
On the lattice however, these results are not reproduced exactly. Topological 
zero modes are shifted away from zero and averages of the lattice 
$\Gamma_5$ operator can differ significantly from the numbers 
given above \cite{VINK}. By combining the two pieces of information 
($\lambda$ and $r$), it is however often possible to decide on the topological 
or non-topological character of a given mode. This can be done for example by 
drawing a plot of $|<n|\gamma_5|n>|$ versus $|\lambda_n|$. Diagrams of this 
kind were considered by Hands and Teper in zero temperature QCD \cite{HANDS}. 
Here, we will focus our attention on the high temperature regime and leave the 
discussion of the transition region for a longer presentation \cite{KLS}. 
\begin{figure}[htb]
\epsfxsize=7.5cm
\epsffile[70 70 550 550]{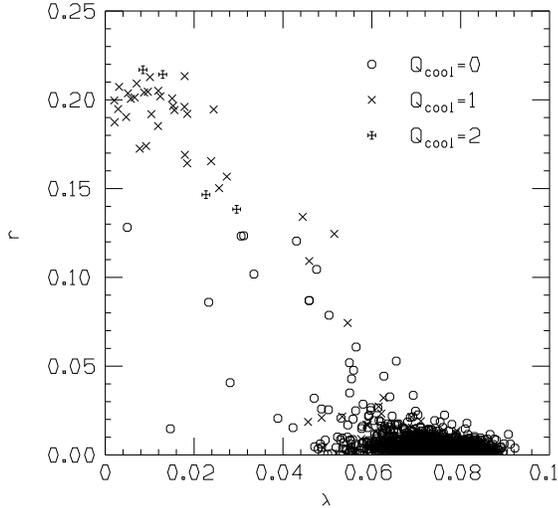}
\vspace{-1.3cm}
\caption{ $(r,\lambda)$ diagram in quenched QCD at $\beta=6.2$ }
\vspace{-0.3cm}
\end{figure}
Our results 
for quenched QCD at $\beta=6.2$ are gathered on figure 2. In order to check 
the consistency of our results with other methods, we also cooled the 
configurations and assigned them an {\it integer} topological charge 
$Q_{cool}$. 
Results obtained on configurations with different $Q_{cool}$ are represented 
by different symbols on figure 2 (see insert). The results behave according 
to expectations; in particular, there is a strong correlation between large 
$r$ and small eigenvalues and topological zero mode candidates can easily be 
identified. Note that they have a value of $r$ around $0.20$ only instead of 
$1$ (This is to be expected since $\Gamma_5$ being a 4-link operator 
picks up a large correction factor already in the mean-field approximation).
By looking at these results configuration by configuration, one can check 
that the number of topological (zero) modes is exactly $4 Q_{cool}$ 
(the 4 being associated with the 4 flavor of staggered fermions). 
The unique configuration with topological charge 2 in our sample, 
for example, gives us 4 low modes in figure 2 
(the existence of 4 other modes with opposite eigenvalue being guaranteed 
by the symmetries of the lattice Dirac operator). Finally, figure 2 also shows 
4 modes with low eigenvalues but small $r$. They all come from one given 
configuration with topological charge 0. An analysis of the local values 
of $<n|\Gamma_5(x)|n>$ reveals that this configuration contains an $I-\bar{I}$ 
pair and that the 4 fermionic modes are delocalized over the 2 topological 
objects. In the continuum, this would give rise to a small but non-zero 
eigenvalue. On the lattice, the 4 copies have different eigenvalues and 
different amount of delocalization because of flavor symmetry breaking. 
In conclusion of this analysis, we see that in quenched QCD at $\beta=6.2$ 
the disconnected pseudoscalar correlator is dominated by the contribution 
from topological zero modes (most of them with topological charge $\pm 1$). 
As the temperature is lowered, the number of small but non-exact zero modes 
increases and the region close to the origin fills up in the $(\lambda,r)$ 
diagram. Finally, we should mention that the 
shift in the eigenvalue of topological modes (extending at least to 
the interval [-0.02,0.02] in figure 2) implies that we should be very 
careful in attempting an extrapolation to the chiral limit. Using masses 
lower than 0.02 for example would lead to a significant underestimate of 
the quantities measured. It is also important to realize that the shift of 
topological modes is an ultraviolet effect (which could for example be 
corrected by the use of an improved or perfect action) and is therefore 
completely different from the depletion of eigenvalues of order $1/V$ 
commonly observed in measurements of $<\bar{\psi}\psi>$ (which is a finite 
size effect).

\end{document}